\documentstyle[sprocl,epsfig]{article}

\def\ab{\bar{\alpha}}
\def\be{\begin{equation}}
\def\ee{\end{equation}}
\def\bea{\begin{eqnarray}}
\def\eea{\end{eqnarray}}

\newcommand{\ltsima}{$\; \buildrel < \over \sim \;$}
\newcommand{\simlt}{\lower.5ex\hbox{\ltsima}}
\newcommand{\gtsima}{$\; \buildrel > \over \sim \;$}
\newcommand{\simgt}{\lower.5ex\hbox{\gtsima}}

\newcommand{\cgs}{${\rm erg~cm}^{-2}~{\rm s}^{-1}$} 
\newcommand{\lum}{\rm erg s$^{-1}$}

\def\lesssim{\mathrel{\hbox{\rlap{\hbox{\lower4pt\hbox{$\sim$}}}\hbox{$<$}}}}
\def\gtrsim{\mathrel{\hbox{\rlap{\hbox{\lower4pt\hbox{$\sim$}}}\hbox{$>$}}}}

\def\arcmin{\hbox{$^\prime$}}

\def\ab1450{$AB_{1450(1+z)}$}

\def\chandra{{\it Chandra\/}}

\def\heao1{{\it HEAO-1\/}}
\def\hst{{\it {\it HST}\/}}

\def \sait #1 #2 {{\em Mem.\ Soc.\ Astron.\ It.\/} {\bf #1}, #2\ }
\def \mess #1 #2 {{\em The Messenger\/} {\bf #1}, #2\ }
\def \astrnach #1 #2 {{\em Astron. Nach.\/} {\bf #1}, #2\ }
\def \aap #1 #2 {{\em Astron. Astrophys.\/} {\bf #1}, #2\ }
\def \aal #1 #2 {{\em Astron. Astrophys. Lett.\/} {\bf #1}, L#2\ }
\def \aar #1 #2 {{\em Astron. Astrophys. Rev.\/} {\bf #1}, #2\ }
\def \aas #1 #2 {{\em Astron. Astrophys. Suppl. Ser.\/} {\bf #1}, #2\ }
\def \aj #1 #2 {{\em Astron. J.\/} {\bf #1}, #2\ }
\def \araa #1 #2 {{\em Ann. Rev. Astron. Astrophys.\/} {\bf #1}, #2\ }
\def \apj #1 #2 {{\em Astrophys. J.\/} {\bf #1}, #2\ }
\def \apjl #1 #2 {{\em Astrophys. J. Lett.\/} {\bf #1}, L#2\ }
\def \apjs #1 #2 {{\em Astrophys. J. Suppl.\/} {\bf #1}, #2\ }
\def \apss #1 #2 {{\em Astrophys. Space Sci.\/} {\bf #1}, #2\ }
\def \asr #1 #2 {{\em Adv. Space Res.\/} {\bf #1}, #2\ }
\def \baic #1 #2 {{\em Bull. Astron. Inst. Czechosl.\/} {\bf #1}, #2\ }
\def \jqsrt #1 #2 {{\em J. Quant. Spectrosc. Radiat. Transfer\/} {\bf #1}, #2\ }
\def \mnras #1 #2 {{\em Mon. Not. R. Astr. Soc.\/} {\bf #1}, #2\ }
\def \mem #1 #2 {{\em Mem. R. Astr. Soc.\/} {\bf #1}, #2\ }
\def \plr #1 #2 {{\em Phys. Lett. Rev.\/} {\bf #1}, #2\ }
\def \pasj #1 #2 {{\em Publ. Astron. Soc. Japan\/} {\bf #1}, #2\ }
\def \pasp #1 #2 {{\em Publ. Astr. Soc. Pacific\/} {\bf #1}, #2\ }
\def \nat #1 #2 {{\em Nature\/} {\bf #1}, #2\ }
\def \etal {{\it et~al.}}

\begin{document}

\title{ON THE NATURE OF EXTREMELY RED OBJECTS: 
THE 2~Ms CHANDRA DEEP FIELD-NORTH SURVEY RESULTS}

\author{C. VIGNALI, D.~M. ALEXANDER, F.~E. BAUER, W.~N. BRANDT, \\
A.~E. HORNSCHEMEIER, G.~P. GARMIRE, D.~P. SCHNEIDER}

\address{Department of Astronomy \& Astrophysics, 
The Pennsylvania State University, \\
525 Davey Laboratory, University Park, PA 16802, USA \\ 
E-mail: chris, davo, fbauer, niel, annh, garmire, and dps@astro.psu.edu}


\maketitle
\abstracts{
Understanding the nature of Extremely~Red~Objects [EROs; ($I-K$)$\ge$4] is one of the most 
challenging issues in observational cosmology. Here we report on the 
\hbox{X-ray} constraints provided by 
the 2~Ms \chandra\ Deep Field-North Survey \hbox{(CDF-N)}. 
X-ray emission has been detected from 11 out of 36 EROs ($\approx$~30\%). 
Five of these have hard X-ray emission and 
appear to be obscured AGNs, while non-AGN emission 
(star formation or normal elliptical galaxy emission) is likely to be 
the dominant source of X-rays from the soft \hbox{X-ray} sources 
detected at the faintest X-ray flux levels. 
}

\section{Introduction}

EROs were serendipitously discovered more than a decade ago{\,}\cite{er88} 
and have optical/near-IR colors consistent with those 
expected for passively evolving 
elliptical galaxies and dust-enshrouded star-forming galaxies{\,}\cite{pm00} 
at $z$$\simgt$1. 
Evidence for elliptical galaxies within the ERO population 
has been found from \nolinebreak optical 
spectroscopic and morphological studies,{\,}\cite{c02,m00} 
while submm-IR observations \nolinebreak have provided clear 
examples of dust-enshrouded galaxies.{\,}\cite{c98,a01}
However, until \nolinebreak recently the fraction of the ERO population hosting AGNs 
was unknown. 
Arguably, the clearest evidence for AGN activity is found from 
X-ray observations.{\,}\cite{h01}
\nolinebreak Here we \nolinebreak provide an update on the 1~Ms CDF-N results{\,}\cite{a02} 
using the 2~Ms \nolinebreak \chandra\ exposure, yielding 
the tightest X-ray constraints on the ERO 
population \nolinebreak to \nolinebreak date.

\section{The samples}

The 2~Ms CDF-N is comprised of 20 observations taken over 27 months and 
reaches an on-axis \hbox{0.5--2~keV} (2--8~keV) flux limit of 
$\approx$~1.5$\times10^{-17}$~\cgs \linebreak 
($\approx$~1$\times10^{-16}$~\cgs).{\,}\cite{a03} 
The region investigated in this study covers \hbox{8.4\arcmin$\times$8.4\arcmin\ }centered 
on the Hubble Deep Field-North (HDF-N). 
Two samples are defined in this study: 
1) {\sf Moderate-depth sample [$K<20.1$]}, including 29 EROs 
(10 with X-ray detections; $\approx$~34\%); 
2) {\sf Deep Sample [$K<22$]}, comprising 9 EROs in 
the HDF-N only (3 with X-ray detections; $\approx$~33\%), 2 of these 
EROs are in common with the moderate-depth sample. 
Seven EROs are detected in \hbox{X-rays} by {\sc wavdetect} using a false-positive probability 
threshold of $10^{-7}$; \nolinebreak for \newpage \noindent the others,  
thresholds of $10^{-5}$ (2 EROs) and $10^{-4}$ (2 EROs) were adopted.

\section{Colors and morphologies of X-ray detected EROs}

The $V-I$ versus $I-K$ colors of all EROs are shown in Fig.~1. 
The overall colors of EROs are consistent with those of $z\approx$~1--2 
early-type spiral and elliptical galaxies.
\begin{figure}[!h]
\vskip -0.5 truecm
\centering
\epsfig{figure=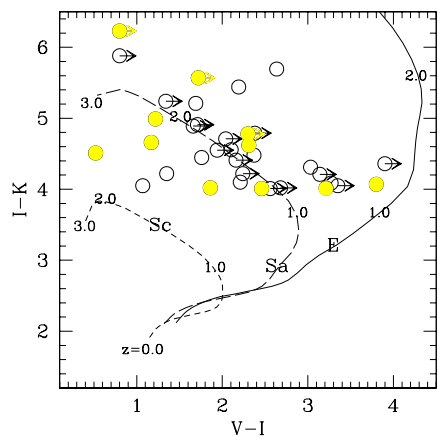,width=0.5\textwidth}
\vskip -0.35cm
\caption{$V-I$ vs. $I-K$ colors for all of the EROs in the two samples. 
The filled circles indicate the X-ray detected EROs. 
Note that the galaxy tracks shown do not include evolutionary effects and 
consequently present redder $V-I$ colors than those expected.}
\label{fig1}
\end{figure}
The X-ray detected EROs 
(filled circles) have the same spread in \vskip -0.3cm \noindent
colors as the X-ray undetected ones, 
suggesting that the X-ray emission is not dependent on the host galaxy type. 
Two examples of X-ray detected EROs in the HDF-N 
are shown in Fig.~2, demonstrating the diversity of the ERO population. 
In the left panel the ERO (the left-most galaxy) 
is a $z$=1.52 emission-line galaxy with a spiral/disturbed morphology, 
while in the right panel the ERO (at $z$=1.32 photometric) 
has an elliptical morphology.
\begin{figure}[!h]
\vskip 3.7cm
\centering
\epsfig{figure=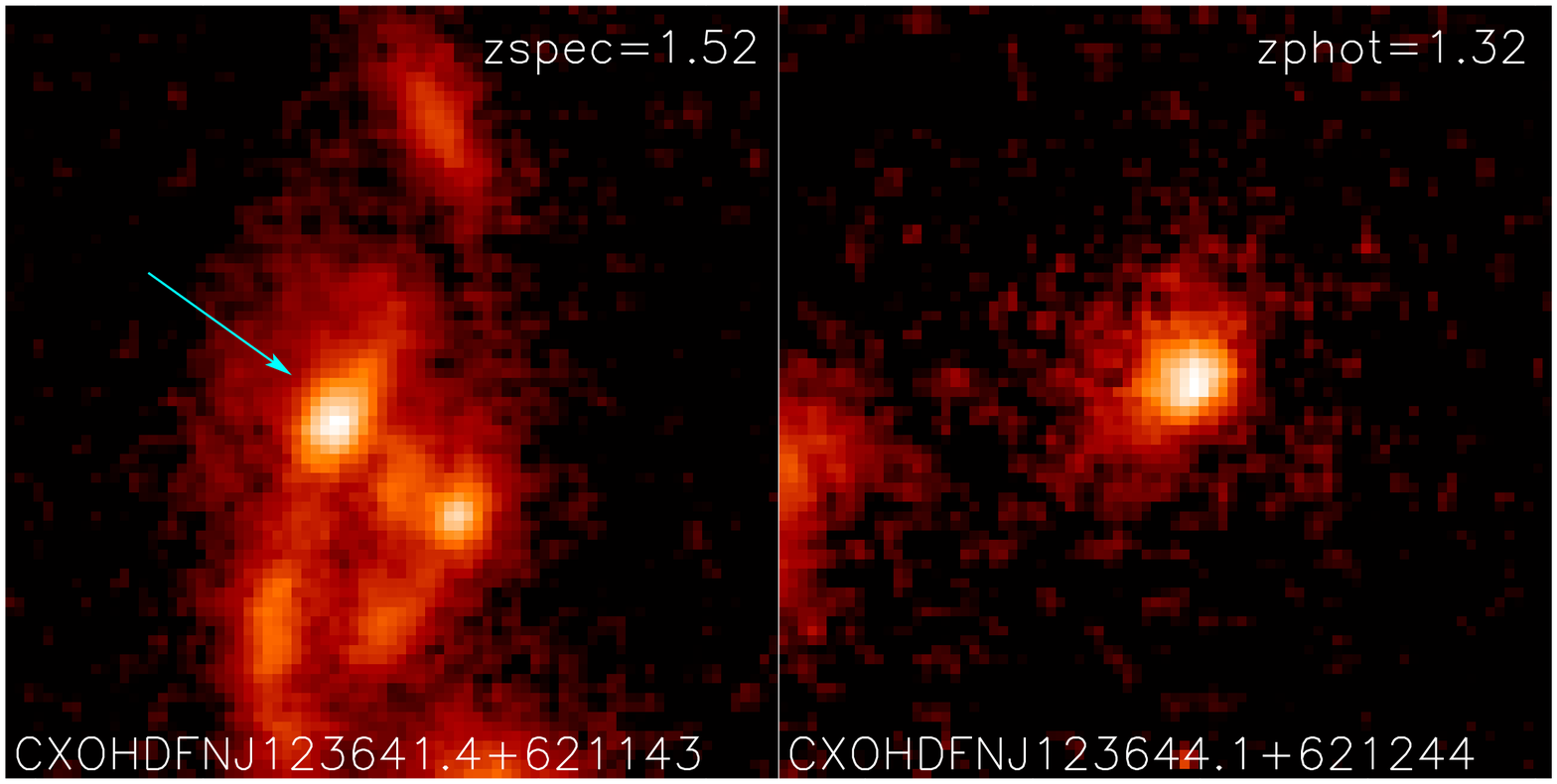,width=0.6\textwidth,bbllx=0pt,bblly=75pt,bburx=680pt,bbury=430pt}
\vskip -0.4cm
\caption{Two examples of X-ray detected EROs in the HDF-N 
(\hst\ $F814W$-filter images).}
\label{fig2}
\end{figure}

\section{X-ray constraints on the nature of EROs}

The effective X-ray photon index ($\Gamma$) of a source can provide a basic 
constraint on its nature (see Fig.~3). 
Obscured AGNs have considerably flatter $\Gamma$ 
than the canonical $\Gamma\approx$~2.0 photon index of 
unobscured AGNs{\,}\cite{g00} due to photo-electric absorption of the 
X-ray emission, and they often have 
\hbox{X-ray} luminosities larger than 10$^{42}$~\lum. 
On the other hand, star-forming galaxies often have \hbox{X-ray} emission 
consistent with that of $\Gamma\approx$~2.0 
power-law emission,{\,}\cite{kft92} 
and very few starburst galaxies have X-ray luminosities in excess of 
\hbox{$\approx$~10$^{42}$~\lum}. 
Passive ellipticals have often X-ray emission consistent with $\Gamma$=1.6--1.8 
and X-ray luminosities $\simlt$10$^{41}$~\lum.{\,}\cite{p99}
%
\begin{figure}[!h]
\vskip -0.55 truecm
\centering
\epsfig{figure=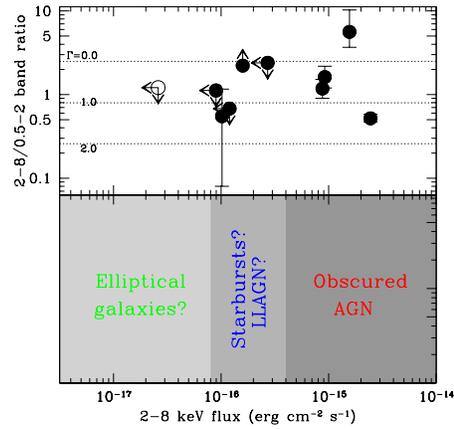,width=0.5\textwidth}
\vskip -0.45 truecm
\caption{Band ratios (i.e., the ratio of the \hbox{2--8~keV} 
to \hbox{0.5--2~keV} count rates) 
vs. 2--8~keV fluxes for the 9 EROs with X-ray constraints (filled circles). 
The open circle shows the stacking analysis result for the X-ray undetected EROs. 
Effective $\Gamma$ are also indicated.} 
\label{fig3}
\end{figure}
\vglue -0.30cm
The X-ray detected EROs with $\Gamma$$<$1.0 are likely to host obscured AGNs, 
and their X-ray luminosities (few $\times$~10$^{42}$~\lum at $z$=1) 
agree with this interpretation. Also, the X-ray brightest ERO is likely to be 
an AGN based on its X-ray luminosity ($\approx$~10$^{43}$~\lum\ at $z$=1). 
Therefore at least 5 of the X-ray detected EROs 
are likely to host AGNs ($\approx$~50\% of the X-ray detected EROs and 
$\approx$~14\% of the ERO population). 
The average stacked X-ray spectral slope of the other 4 
EROs with X-ray constraints ($\Gamma$$>$1.0), their X-ray luminosities ($<$$10^{42}$~\lum at $z$=1), 
and multi-wavelength constraints 
suggest that they could be starburst galaxies{\,}\cite{a02} 
or low-luminosity AGNs. 
Some of these sources also tend to have irregular optical morphologies and 
extremely red colors [i.e., ($I-K$)$>$5]. 
Stacking analysis of the X-ray undetected EROs provides an \hbox{X-ray} detection in 
the 0.5--2~keV band (open circle in Fig.~3) 
consistent with emission from $L_{\star}$ elliptical galaxies. 
Further constraints on the nature of X-ray detected EROs come from X-ray 
spectral analysis of the four EROs with more than 80 counts (see Fig.~4 
for one example). 
\begin{figure}[!h]
\vskip -0.2 truecm
\centering
\epsfig{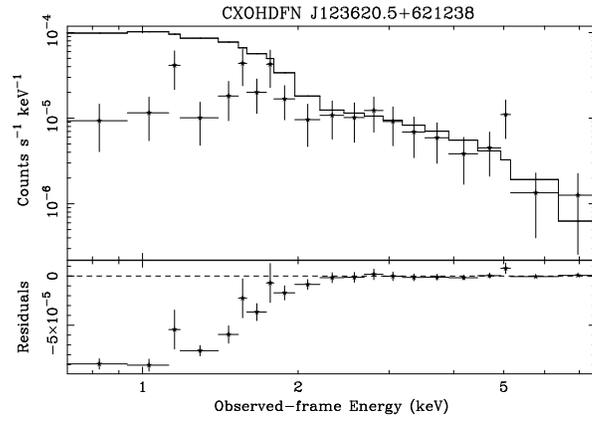}
\vskip -0.3truecm
\caption{CXOHDFN~J123620.5$+$621238 (binned to 5 counts/bin for presentation purposes) 
fitted with an unabsorbed $\Gamma$=2 power law. 
A deficit of counts below 2~keV is clearly present. 
This source has $\approx$~110 counts in the \hbox{0.5--8~keV} band.} 
\label{fig4}
\end{figure}
When modeled with $\Gamma$=2 power laws (using the Cash statistic), 
all four appear to be absorbed by column densities 
of $N_{\rm H}\approx$~(0.1--2.0)$\times10^{23}$~cm$^{-2}$ 
(at $z$=1). There is no strong evidence for iron K$\alpha$ emission lines. 

%
%
%

\end{document}